# On a heuristic point regarding the origin of tachyons and a revamped definition for them

Aasis Vinayak.P.G.

Varadakshina, Perayam, Mulavana.P.O.,

Kollam, Kerala - 691503, India

Abstract

As it follows from the classical analysis of the data, the paper is intended to solve one of the main problems in the physics of tachyons – they are having imaginary mass. Further we use QM to deal with the matters concerning the 'particle'. The paper also considers the physical viability of tachyons in real world. The paper also considers the natural occurrence and persistence of tachyons.

*Key words: Tachyon, matter wave, FTL.*

## 1. Introduction

In the introductory part I consider only some present notions regarding the tachyons and FTL phenomenon. Imaginary mass is a bizarre theoretical concept that comes from taking the square root of a negative number; in this case, it roughly means that a particle's mass is only "physically meaningful" at speeds greater than light. Tachyons have never been found in experiments as real particles traveling through the vacuum, but we predict theoretically that tachyon-like objects exist. In case associated with faster-than-light 'quasiparticles' moving through laser-like media similar are the aspects. There are strong scientific reasons to believe that they really exist, because Maxwell's equations, when coupled to inverted atomic media, lead inexorably to tachyon-like solutions. Even if we discover such a particle, say tachyon, its mass 'may be imaginary' and it is a troubling question in mathematics.

Electronic mail: aasisvinayak@gmail.com

It is note worthy to mention a work here - Quantum optical effects can produce a different kind of 'faster than light' effect [1]. There are actually two different kinds of 'faster-than-light' effects that we have found in quantum optics experiments.

First, we have discovered that photons which tunnel through a quantum barrier can apparently travel faster than light [2]. Because of the uncertainty principle [3], the photon has a small but very real chance of appearing suddenly on the far side of the barrier, through a quantum effect (the 'tunnel effect') which would seem impossible according to classical physics. The tunnel effect is so fast that it seems to occur faster than light.

Now consider the aspects concerning the some details of a quasi particle say a phonon as they may exhibit similar properties.(But it is not an essential part for this paper, here it is considered to discuss the aspect vividly) The formalism that we will develop for the model is readily generalisable to two and three dimensions. The Hamiltonian for this system is (thus for phonons can be denoted as)

$$\mathbf{H} = \sum_{i=1}^{N} \frac{p_i^2}{2m} + \frac{1}{2} m\omega^2 \sum_{\{ij\}(nn)} (x_i - x_j)^2 \tag{1.1}$$

where $m$ is the mass of each atom, and $x_i$ and $p_i$ are the position and momentum operators for the $i$th atom. A discussion of similar Hamiltonians may be found in the article on the quantum harmonic oscillator. It is straightforward, though tedious, to generalize the above to a three-dimensional lattice. One finds that the wave number $k$ is replaced by a three-dimensional wave vector **k**. Furthermore, each **k** is now associated with *three* normal coordinates. The Hamiltonian has the form

$$\mathbf{H} = \sum_{k} \sum_{s=1}^{3} \hbar \omega_{k,s} \left( a_{k,s}^{\dagger} a_{k,s} + 1/2 \right) \tag{1.2}$$

The new indices $s = 1, 2, 3$ label the polarization of the phonons.

By inverting the discrete Fourier transforms to express the $Q$'s in terms of the $x$'s and the $\Pi$'s in terms of the $p$'s, and using the canonical commutation relations between the $x$'s and $p$'s, we can show that

$$[Q_k, \Pi_{k'}] = i\hbar \delta_{kk'} \quad ; \quad [Q_k, Q_{k'}] = [\Pi_k, \Pi_{k'}] = 0 \tag{1.3}$$

Second, we have found an effect related to the famous Einstein-Podolsky-Rosen phenomenon [4], in which two distantly separated photons can apparently influence one another's' behaviors at two distantly separated detectors [5]. This effect was first predicted theoretically by Prof. J. D. Franson of Johns Hopkins University. We have found experimentally that twin photons emitted from a common source (a down-conversion crystal) behave in a correlated fashion when they arrive at two distant interferometers. This phenomenon can be described as a 'faster-than-light influence' of one photon upon its twin. Because of the intrinsic randomness of quantum phenomena, however, one cannot control whether a given photon tunnels or not, nor can one control whether a given photon is transmitted or not at the final beam splitter. Hence it is impossible to send true signals in faster-than-light communications. In the case associated with the conditions for tachyon condensation and using string theory (with help of strings and branes) we are arriving again at the same bizarre situation that the tachyons are having imaginary rest mass, but of course at a more convincing way that they may exist. Hence there are so many reasons to believe that tachyon and similar particles may exist.

Thus for a tachyon we can say that, if $m_o$ is the rest mass of the particle then the relative mass of the particle is $m_v$ (if the velocity is v) which is related to the rest mass as [6] (It should be noted that some equations [esp. of Quantum Mechanics] are given as standard ones – non relativistic- which can be transformed to applicable ones-here it is given as such in order to avoid confusion).

$$m_v = \frac{m_0}{\sqrt{1-\frac{v^2}{c^2}}} \qquad (1.4)$$

$$m_v = \frac{m_0}{\sqrt{\frac{c^2-v^2}{c^2}}} \qquad (1.5)$$

Here the velocity of the particle is greater than the velocity of light; hence the factor inside the square root will be imaginary.

$$m_v = \frac{m_0}{\sqrt{\frac{(v^2-c^2)\times -1}{c^2}}} \qquad (1.6)$$

$$m_v = \frac{m_0}{\sqrt{-1}\times\sqrt{\frac{(v^2-c^2)}{c^2}}} \qquad (1.7)$$

If we put $\sqrt{\frac{(v^2-c^2)}{c^2}} = n$, an integer;

Then,

$$m_v = \frac{m_0}{i \times n} \quad (1.8)$$

-which an imaginary number- From (1.8) if we do like this the corresponding momentum and energy will also be similar. To avoid this, we may say that the rest mass of the tachyon is imaginary. Thus 'its' momentum and energy will be real. Hence the tachyon will have imaginary rest mass; that is bizarre in mathematics and in physical world. To solve this problem the complex form has to be solved.

But we must now consider an important aspect with respect to these particles. When it is moving with such a high velocity as specified earlier it will have more wave nature. To make it more convincing now let us assume that the velocity $v$ is very high, say is near to infinity, so that even if the real rest mass of the particle is very high its relative mass at $v$ will be small. Rather we can say that it only a matter wave at this 'turning point'. We have to consider the probability of finding the tachyon at a point. Yet the total change of finding over an area wide enough to contain it is 100% by the relation in Quantum Mechanics. To find the tachyon in specified region we can use the integral of the probability.

$$\frac{\partial}{\partial t}\int_\Omega P(r,t)d^3r = \frac{ih}{4\pi m}\int_\Omega \nabla \cdot [\psi^* \nabla \psi - (\nabla \psi^*)\psi]d^3r \quad (1.9)$$

*P(r,t)* here shows the position probability density. To make the wave packet more distinct we took this by computing the time derivative of the integral of *P* over any fixed volume *Ω*

$$\frac{\partial}{\partial t}\int_\Omega P(r,t)d^3r = \frac{ih}{4\pi m}\int_A [\psi^* \nabla \psi - (\nabla \psi^*)\psi]_n dA \quad (1.10)$$

This integral is obtained by the partial integration (using Green's theorem) and here *A* is the bounding surface of the region of integration and [ ]$_n$ denotes the component of the vector in the brackets in the direction of the outward normal to the surface element *dA*.[ If there a force derivable from real potential energy it could also be included (but here the case is not so)]. Hence we need to apply the advanced form.

Thus by considering it as a matter wave we could write (based on assumption regarding *v*)

$$\lambda = \frac{h}{m \times v} \quad (1.11)$$

Or in very specific way (1.11) can be written as

$$\lambda_v = \frac{h}{m_v \times v} \tag{1.11.a}$$

, if the velocity of the tachyon is $v$ and its relative mass $m_v$ and the wavelength (de broglie's wave length of the matter wave) of wave is $\lambda_v$ [7]

$$m_v = \frac{h \times v_v}{v^2} \tag{1.12}$$

here the frequency is $v_v$). And the position probability of such a particle an be generalised simply as

$$P(r,t) = \psi^*(r,t)\psi(r,t) = |\psi(r,t)|^2 \tag{1.13}$$

We have specified (1.13) clearly as (1.9) and (1.10). Taking all the materials let us move on to the next part of the paper.

## 2. A Review

*Now let us assume the rest mass of tachyon, $m_0$, is not imaginary* and is travelling with velocity near to the velocity of light, say u. (Here it should be noted that the tachyon is not travelling in the way specified, but the corresponding rest mass, $m_o$, is made to travel like this).

Here when the particle having rest mass $m_o$ is moving with a velocity, say $u$, very close to the velocity of light then it will have definitely wave nature –it has this even if the velocity is small-(matter wave and the nature increases with increase in velocity) – whether it moves harmonically or as complex wave function (but here the magnitude of the wave length of the matter wave is not a determining factor, that we will see later) we can find the relative mass of the particle when it has acquired this velocity $u$ which is presumed to be very close to the velocity of light c [6].

$$m_u = \frac{m_0}{\sqrt{1-\frac{u^2}{c^2}}} \tag{2.1}$$

$$m_0 = m_u \times \sqrt{1-\frac{u^2}{c^2}} \tag{2.2}$$

From (1.12) we can state as [7]

$$m_0 = \frac{h \times v_u}{u^2} \times \sqrt{1 - \frac{u^2}{c^2}} \tag{2.3}$$

And also as it is well elucidated that $\omega_u = 2\pi v_u = \sqrt{\frac{k_u}{m_u}}$ (2.4)

, where $\omega_u$ its angular velocity and $k_u$ is is its corresponding force constant (which will come to the equation when we consider the whole particle as a wave packet esp. from the wave equations concerning simple harmonics motion and transformations; and later on applications to harmonic and complex wave functions, as it is merely the linear superposition of this and also here the 'matter' which is moving is responsible for this )

$$m_0 = \frac{h \times \omega_u}{2\pi u^2} \times \sqrt{1 - \frac{u^2}{c^2}} \tag{2.5}$$

It should be noted that the wave packet has this due to the super position of many waves and here in all the cases the mass is a determining factor(hence $k$ also) which is responsible for the same when acquires some velocity.

$$m_0 = \frac{h \times \sqrt{\frac{k_u}{m_u}}}{2\pi u^2} \times \sqrt{1 - \frac{u^2}{c^2}} \tag{2.6}$$

Again using the relation Eq.(2.4) we can say that

$$m_0 = \frac{h \times \sqrt{\frac{k_u \times u^2}{h \times v_u}}}{2\pi u^2} \times \sqrt{1 - \frac{u^2}{c^2}} \tag{2.7}$$

In order to get the expectation value of the position vector of a particle, we write the expectation value for **r** as we know the fact that

$$\langle r \rangle = \int r P(r,t) d^3r = \int \psi^*(r,t) r \psi(r,t) d^3r \tag{2.8}$$

As Eq.(2.8) is equivalent to the equations (2.9), (2.10) and (2.11) (in order to find the 'particle')

$$\langle x \rangle = \int \psi^* x \psi d^3r \tag{2.9}$$

$$\langle y \rangle = \int \psi^* y \psi d^3 r \tag{2.10}$$

$$\langle z \rangle = \int \psi^* z \psi d^3 r \tag{2.11}$$

where ψ is normalised. The expectation value is a function only of time, since ψ and P depend on t and space coordinates have been integrated out.

Now when a tachyon is moving with a velocity *v* (which is even greater than the velocity of light) it will have more wave nature than that of particle nature – as elucidated above. Thus in an over simplified sense it is no longer a particle is only a matter wave (as it has only probability).

Now let us relate the rest mass of the tachyon and its relative mass.

Putting the value from (2.7) of $m_o$ in (1.5) we will get

$$m_0 = \frac{\frac{h \times \sqrt{\frac{k_u \times u^2}{h \times v_u}}}{2\pi u^2} \times \sqrt{1 - \frac{u^2}{c^2}}}{\sqrt{-1} \times \sqrt{\frac{(v^2 - c^2)}{c^2}}} \tag{2.12}$$

It can be simplified as

$$m_v = \frac{1}{2\pi} \times \sqrt{\frac{hk_u}{-v_u}} \times \sqrt{\frac{(c^2 - u^2)}{(v^2 - c^2)u^2}} \tag{2.13}$$

We need to keep in mind that the probability of finding the particle somewhere in the region must be unity, so that Eq.(1.13) implies that the wave function is normalised:

$$\int |\psi(r,t)|^2 d^3 r = 1 \tag{2.14}$$

Also, as we know the energy eigenvalues are only possible results of precise measurement of the total energy and that the probability of finding a particular value *E* when the particle is described by the wave function ψ(r) is proportional to $|A_E|^2$ - from standard results. It is easily seen that the proportionality factor is unity because if we put for the energy probability function

$$P(E) = |A_E|^2$$

We see that *P(E)* sums to unity

$$\sum_E P(E) = \iint \psi^*(r')\psi(r)[\sum_E u_E^*(r)u_E(r')]d^3rd^3r' = 1 \qquad (2.23)$$

since ψ is normalized as (as per standard results)

$$\int \sum_E u_E^*(r')u_E(r)]d^3r' = 1$$

So we can also compute the average or expectation value of the energy from the probability function:

$$\langle E \rangle = \sum_E EP(E) = \sum_E \int Eu_E^*(r)\psi(r)d^3r \int u_E(r')\psi^*(r')d^3r' \qquad (2.24)$$

Keeping this in mind we can proceed further.
As we know the fact, from Eq. (2.4),

$$v_u = \frac{\omega}{2\pi} \qquad (2.15)$$

And also

$$k_u = \omega^2 \times m_u \qquad (2.16)$$

Thus Eq.(2.13) will become

$$m_v = \frac{1}{2\pi} \times \sqrt{\frac{h\omega_u^2 \times m_u \times 2\pi}{-\omega}} \times \sqrt{\frac{(c^2 - u^2)}{(v^2 - c^2)u^2}} \qquad (2.17)$$

Hence,

$$m_v = \frac{1}{2\pi} \times \sqrt{\frac{-1 \times h\omega_u \times m_u \times 2\pi \times (c^2 - u^2)}{(v^2 - c^2)u^2}} \qquad (2.18)$$

which can be simplified as,

$$m_v = \sqrt{\frac{-1 \times h\omega_u \times m_u \times (c^2 - u^2)}{(v^2 - c^2)u^2 \times 2\pi}} \qquad (2.19)$$

Again applying Eq. (2.15)

$$m_v = \sqrt{\frac{-1 \times hv_u \times m_u \times (c^2 - u^2)}{(v^2 - c^2)u^2}} \qquad (2.20)$$

$$m_v = \sqrt{\frac{-1 \times h \times u \times m_u \times (c^2 - u^2)}{\lambda_u \times (v^2 - c^2)u^2}} \qquad (2.21)$$

$$m_v = \sqrt{\frac{h \times m_u \times (c^2 - u^2)}{-1 \times u \times \lambda_u \times (v^2 - c^2)}} \tag{2.22}$$

Here it means that means that the velocity will be reversed. (It should be kept in mind that *u* is the velocity and v is the frequency and here the direction is only reversing the magnitude may change with respect to the velocity of the tachyon *v* as there are other terms long with, in the equation.).(It should be noted that the phase velocity and group velocity-which equal to *v* has to be taken into account while dealing with these aspects). If we consider this direction, opposite direction, as positive, then the relation (2.21) becomes (applying sign convention)

$$m_v = \sqrt{\frac{h \times m_u \times (c^2 - u^2)}{u \times \lambda_u \times (v^2 - c^2)}} \tag{2.23}$$

This implies that the direction of propagation of the matter wave of the tachyon will be opposite to that when the same rest mass travels with the velocity less than that of light. (Here we need to look at the fourth part of the part – '*discussion*')

## 3. In 'Classical terms'

From the equation (2.20) .As
$$\lambda_u = \frac{h}{m_u u} \tag{3.1}$$

Also,
$$\lambda_u \times \lambda_u \times m_u \times v_u = h \tag{3.2}$$
Thus,
$$m_u \times v_u = \frac{h}{(\lambda_u)^2} \tag{3.3}$$
So we can say that [from (2.20)], by considering *the direction as positive*

$$m_v = \sqrt{\frac{h^2 \times (c^2 - u^2)}{(\lambda_u)^2 \times (v^2 - c^2) u^2}} \tag{3.4}$$

Or,

$$\lambda_u = \sqrt{\frac{h^2 \times (c^2 - u^2)}{(m_v)^2 \times (v^2 - c^2)u^2}} \qquad (3.5)$$

Now when we have the relations the following three relations

$$\langle p_x \rangle = -i\frac{h}{2\pi}\int \psi^* \frac{\partial \psi}{\partial x} d^3r \qquad (3.6)$$

$$\langle p_y \rangle = -i\frac{h}{2\pi}\int \psi^* \frac{\partial \psi}{\partial y} d^3r \qquad (3.7)$$

$$\langle p_z \rangle = -i\frac{h}{2\pi}\int \psi^* \frac{\partial \psi}{\partial z} d^3r \qquad (3.8)$$

We could measure the value of momentum p There will be some uncertainty as detailed above –and will also be considered below) and $m_v$ from the experiment (if we could discover one). Now if we consider a velocity u arbitrarily such that it is near to the velocity of light or so, we could measure the de broglie's wave length of the mass at that velocity (and the rest mass of both remaining constant), using the equation (3.5) and hence the corresponding mass can also be found using the basic relation

$$m_u = \frac{h}{\lambda_u \times u} \qquad (3.9)$$

Thus the corresponding rest mass of the particle can be found out

$$m_0 = m_u \times \sqrt{1 - \frac{u^2}{c^2}} \qquad (3.10)$$

Hence in effect the rest mass of the tachyon can be found out. This has to be substantiated in the classical one also.

If the expectation value is defined in the general case with the operator acting on ψ and multiplied on the left by ψ*, therefore we will obtain the values for example

$$\langle E \rangle = \int \psi^* i\frac{h}{2\pi} \frac{\partial \psi}{\partial t} d^3r \qquad \langle p \rangle = \int \psi^* (-i\frac{h}{2\pi})\nabla \psi d^3r \qquad (3.11)$$

(3.10) and (3.11) are concerning <E> and <p>, we could also theoretically predict about the relative mass of the tachyon. (This is by taking the direction of propagation to be negative, as per earlier derived results)

So we can say using the relation connecting mass and the frequency we can say from (2.17) that

$$m_v = \frac{1}{2\pi} \times \sqrt{\frac{h \times \omega_u^2 \times m_u (c^2 - u^2)}{-v_u \times (v^2 - c^2)u^2}} \qquad (3.12)$$

$$m_v = \frac{1}{2\pi} \times \sqrt{\frac{h \times 4\pi^2 \times m_u(c^2 - u^2) \times v_u^2}{-v_u \times (v^2 - c^2)u^2}} \tag{3.13}$$

$$m_v = \sqrt{\frac{h \times m_u(c^2 - u^2) \times -v_u}{(v^2 - c^2)u^2}} \tag{3.14}$$

As we know the fact

$$m_v = \frac{h}{\lambda_v \times v} \tag{3.15}$$

This implies that

$$\frac{h}{\lambda_v \times v} = \sqrt{\frac{h \times m_u(c^2 - u^2) \times -v_u}{(v^2 - c^2)u^2}} \tag{3.16}$$

$$\frac{h^2}{(\lambda_v \times v)^2} = \frac{h \times m_u(c^2 - u^2) \times -v_u}{(v^2 - c^2)u^2} \tag{3.17}$$

As

$$\frac{h}{\lambda_v \times v} = \frac{h \times v_v}{v^2} \tag{3.18}$$

Hence

$$\frac{h^2 \times v_v^2}{v^4} = \frac{h \times m_u(c^2 - u^2) \times -v_u}{(v^2 - c^2)u^2} \tag{3.19}$$

$$\frac{v^4}{h \times v_v^2} = \frac{v^2 u^2 - c^2 u^2}{m_u \times -v_u \times (c^2 - u^2)} \tag{3.20}$$

$$\frac{m_u \times -v_u \times (c^2 - u^2)}{h \times v_v^2} = \frac{u^2 \times v^2 - c^2 \times u^2}{v^4} \tag{3.21}$$

$$[m_u \times -v_u \times (c^2 - u^2)] \times v^4 = [h \times v_v^2 u^2]v^2 - [h \times v_v^2 \times c^2 \times u^2] \tag{3.22}$$

As this is of the form $ax^2 + bx + c = 0$ we can find the value of $v^2$ just like we find the value of x in the stand form, by applying the relation,

$$x = \frac{-b \pm \sqrt{b^2 - 4ac}}{2a} \tag{3.23}$$

Thus by applying (3.23)

$$v^2 = \frac{[h \times v_v^2 \times u^2] \pm \sqrt{[h \times v_v^2 u^2]^2 + 4m_u \times v_u \times (c^2 - u^2)[h \times v_v^2 \times c^2 \times u^2]}}{-2[[m_u \times v_u \times (c^2 - u^2)]]} \tag{3.24}$$

This is by *the convention* also we know the fact now that 'the factor inside the root' in the classical equation is not negative but it has to be positive for a tachyon also as per the earlier derivation ( as we have proved that it will have real rest mass and not imaginary). Then the numerator of the above equation should be positive as the denominator is negative. Thus by analysis the equation, if we take the positive value of the portion inside the square root [as here the portion inside the square root the greater than that is out side (in the numerator)] the part ($v^2$) will be negative as a whole. Though the equations that we get mathematically are

$$v^2 = \frac{[h \times v_v^2 \times u^2] + \sqrt{[h \times v_v^2 u^2]^2 + 4m_u \times v_u \times (c^2 - u^2)}}{-2[m_u \times v_u \times (c^2 - u^2)]} \tag{3.25}$$

and

$$v^2 = \frac{[h \times v_v^2 \times u^2] - \sqrt{[h \times v_v^2 u^2]^2 + 4m_u \times v_u \times (c^2 - u^2)}}{-2[m_u \times v_u \times (c^2 - u^2)]} \tag{3.26}$$

we cannot take both of them.

We have the classical equation $m_v = \dfrac{m_0}{\sqrt{1 - \dfrac{v^2}{c^2}}}$ \hfill (3.27)

It is clear that, for tachyon, only if $\dfrac{v^2}{c^2}$ becomes negative the portion inside the root in Eq.(3.27) will be positive. Thus by analysing the equations (3.25) and (3.26), we find that (3.25) will only give the desired result [as there is a probability that $\dfrac{v^2}{c^2}$ would become positive in the case of (3.26)].

Now substituting the value of $v^2$ from (3.25) we get

$$m_v = \frac{m_0}{\sqrt{1 + \dfrac{[h \times v_v^2 \times u^2] + \sqrt{[h \times v_v^2 u^2]^2 + 4m_u \times v_u \times (c^2 - u^2)}}{2[m_u \times v_u \times (c^2 - u^2)]}}} \tag{3.28}$$

$$m_v = \frac{m_0}{\sqrt{c^2 + \dfrac{[h \times v_v^2 \times u^2] + \sqrt{[h \times v_v^2 u^2]^2 + 4m_u \times v_u \times (c^2 - u^2)}}{2[m_u \times v_u \times (c^2 - u^2)]}}} \tag{3.29}$$

This one is only a mathematically derived result, but it has more physically viable. There are some other views like the one that we may take the fourth power and later taking the forth root. This may also proved mathematically *but not physically* (the mode is very speculative).

From this newly derived result [not (3.29)] it is quite clear that the particle will have only real mass (both relative and at rest). The reason for which why we need to make a substitution in the classical equation is that in the classical one we have the factor $v^2$ Hence classically also we could get the derived result. By applying the principles of Quantum Mechanics we have seen that the direction is just opposite. So when $v^2$ is there it is nullified by the square of the velocity. This case is applicable for other particles also, but in this case the negative sign persists.

## 4. Discussion

The simplified equation (2.22) will tell us why in classical pathway the mass is imaginary as here when u is negative hence it will give an imaginary number mathematically. But physically it is as elucidated above. Hence it is easy that we can get the exact corresponding value of $\lambda_u$ by putting the values in the derived equation (3.5).

Thus we have seen that, as special relativity restricts, the fact that a sub luminal particle cannot be accelerated to a hyper luminal velocity, the work is occurring like this. A tachyon is a particle that has a velocity greater than that of light at the time of its production itself. From the above-derived result (2.23) we can say that a tachyon will not have an imaginary rest mass. But will have a relative mass and thus related to the mass of the wave packet and it will be propagating in a direction such that the corresponding rest mass when travels with a velocity less than the velocity of light the matter wave propagation will be opposite in opposite direction. We call this point at which the inversion takes place as *Raghava Inversion Point (RIP)*. The significance of this point is detailed below.

There is significance for the above-derived result. Special relativity restricts the acceleration of a sub luminal particle to a hyper luminal velocity. The above derived result also elucidates the same thing but in a different way. If we accelerate a particle like this, its momentum will get reversed. But here there will be violation of the conservation of momentum. Hence it is not possible. *Thus one of the important results of special relativity can also be derived from this proposal*.

 A tachyon may not occur as a charged particle as then there will be violation of conservation of linear momentum when it emits Cerenkov radiation for some specific relative period. Hence we can conclude that a charged particle will never be born with a velocity greater than that of light.

It should also be noted that when the tachyon's velocity is reversed by some means there will not be any change expect in the fact that then the matter wave of the rest mass at velocity less than that of light will also get reversed.

In the case of this wave packet also the uncertainty principle plays some role. As

$$(\Delta x)^2 (\Delta p)^2 \geq \left| \int (\alpha^* \psi^*)(\beta \psi) dx \right|^2 = \left| \int (\psi^* \alpha \beta \, \psi dx \right|^2 \tag{4.1}$$

$$\alpha \equiv x - \langle x \rangle \qquad \beta \equiv p - \langle p \rangle = -i \frac{h}{2\pi} \left( \frac{d}{dx} - \left\langle \frac{d}{dx} \right\rangle \right) \tag{4.2}$$

We can also find $|\psi(r,t)|^2$ by using the relation

$$|\psi(r,t)|^2 = \frac{1}{\sqrt{\{2\pi[(\Delta x)^2 + \frac{h^2 t^2}{16\pi^2 m^2 (\Delta x)^2}]\}}} \exp-\frac{x^2}{2[(\Delta x)^2 + \frac{h^2 t^2}{4\pi^2}/4m^2 (\Delta x)^2]} \tag{4.3}$$

Thus here too there will be some error regarding $x$ and $p$ while measuring them.

It is also possible to decelerate a tachyon to the velocity just greater than the velocity of light. But it cannot be decelerated to the velocity of light or below as then there will be violation of conservation of momentum (which is clear from analysing the derived results).

## References:

1. "Faster than light?" by R. Y. Chiao, P. G. Kwiat, and A. M. Steinberg in *Scientific American*, August 1993

2. "Measurement of the Single-Photon Tunneling Time" by A. M. Steinberg, P. G. Kwiat, and R. Y. Chiao, *Physical Review Letters*, Vol. 71, page 708; 1993

3. W. Heisenberg, *Über den anschaulichen Inhalt der quantentheoretischen Kinematik und Mechanik*, Zeitschrift für Physik, 43 1927, pp 172-198. English translatation: J. A. Wheeler

4. A. Aspect, *Bell's inequality test: more ideal than ever*, Nature **398** 189 (1999). J.S. Bell *On the Einstein-Poldolsky-Rosen paradox*, Physics **1** 195 (1964

5. "High-Visibility Interference in a Bell-Inequality Experiment for Energy and Time," by P. G. Kwiat, A. M. Steinberg, and R. Y. Chiao, *Physical Review A*, Vol. 47, page R2472; 1993

6. Special Relativity Lecture:
   Notes (*http://www.phys.vt.edu/~takeuchi/relativity/notes*) - A standard introduction to special relativity where explanations are based on pictures called space-time diagrams

7. Nave, R., "*Wave-Particle Duality (http://hyperphysics.phy-astr.gsu.edu/hbase/mod1.html)*". HyperPhysics, Quantum Physics